\documentclass[pra,nopacs,twocolumn,nofootinbib]{revtex4}
\usepackage{graphicx}
\usepackage{bm,amsfonts,amsmath,amssymb}
\usepackage{dcolumn}
\usepackage{natbib}
\usepackage
{hyperref}
\hypersetup{
    colorlinks=true,
    linkcolor=blue,
    citecolor=red,
    filecolor=magenta,
    urlcolor=magenta,
}
\newcommand{\Eref}[1]{Eq.~(\ref{#1})}
\newcommand{\Tref}[1]{Table~(\ref{#1})}

\def\vkapp{\varkappa}

\begin{document}

\title{Calculation of thallium hyperfine anomaly}

\author{E. A. Konovalova$^1$}
\author{M. G. Kozlov$^{1,2}$}
\author{Yu. A. Demidov$^{1,2}$}
\author{A. E. Barzakh$^1$}

\affiliation{$^1$ Petersburg Nuclear Physics Institute, Gatchina
188300, Russia}
\affiliation{$^2$ St.~Petersburg Electrotechnical University
``LETI'', Prof. Popov Str. 5, 197376 St.~Petersburg}

\date{\today}

\begin{abstract}
We suggest a method to calculate hyperfine anomaly for many-electron atoms and ions. 
At first, we tested this method by calculating hyperfine anomaly for hydrogen-like 
thallium ion and obtained fairly good agreement with analytical expressions. 
Then we did calculations for the neutral thallium and tested an assumption, that 
the the ratio between the anomalies for $s$ and $p_{1/2}$ states is the same for 
these two systems. Finally, we come up with recommendations about preferable atomic 
states for the precision measurements of the nuclear $g$ factors.
\end{abstract}

\maketitle

\section{Introduction}

In recent years, the precision achieved in resonant ionization spectroscopy experiments coupled with advances in atomic theory has
enabled new atomic physics based tests of nuclear models.
Understanding the occurrence of shape coexistence in atomic nuclei is one of them. 
This phenomenon is associated with existence of both the near-spherical and deformed structures of nuclei for
neutron-deficient isotopes near Z = 82 closed shell. 
The measurements of hyperfine constants and isotope shifts are highly sensitive to 
the changes of nuclear charge and magnetic radii because they depend on
the behavior of the electron wave function near the nucleus. 
The hyperfine structure (HFS) measurements can serve as very
useful tool for understanding of shape coexistence phenomena in atomic nuclei.

Magnetic hyperfine constants $A$ are usually assumed to be proportional to the nuclear magnetic moments. 
However, this is true only for the point-like nucleus. 
For the finite nucleus we need to take into account (i) distribution of the magnetization inside the nucleus
and (ii) dependence of the electron wave function on the nuclear charge radius. 
Former correction is called magnetic (Bohr--Weisskopf~\cite{BW50}) and the latter is called charge correction (Breit-Rosenthal\cite{RB32,CS49}).
Together these corrections are known as hyperfine anomaly \cite{Sha94}. 
Below we discuss how to calculate hyperfine anomaly for many-electron atoms with available atomic packages. 
We use thallium atom as reference system for our calculations, because for this atom there are comprehensive experimental data
\cite{LP56,RLM00,BUWC01,BCUT03,BBFI12} and many theoretical calculations \cite{Sha94,STKAY97,MP95,DFKP98,KPJ01}.

\citet{Sha94} and \citet{STKAY97} found analytical expressions for
the hyperfine anomaly for H-like thallium ion. For the neutral
thallium there is numerical calculation by \citet{MP95}.
Experimentally HFS anomaly is studied much better for neutral Tl
than for respective H-like ion. In the work \cite{GFM00} it has been
suggested, that the ratio between the anomalies for $s$ and
$p_{1/2}$ states remains constant for these two systems. Here we try
to test this assumption.

We use atomic package \cite{KPST15}, which is based on the original
Dirac-Hartree-Fock code \cite{BDT77}. This package is often used to
calculate different atomic properties including hyperfine structure
constants of Tl \cite{DFKP98,KPJ01}, Yb \cite{PRK99a}, Mg
\cite{KPWAT15}, and Pb \cite{PKST16}.

\section{ Theory and methods}

A four component Dirac wavefunction of an electron in a spherically
symmetric atomic potential can be written as \cite{BDT77}:
\begin{align}\label{DiracWF}
 \psi_{n,\vkapp,m}(\bm r) =
 \frac1r
 \left(\!\!
 \begin{array}{c}
 P_{n,\vkapp}(r)
 \Omega_{\vkapp,m}(\omega) \\
 -i\,Q_{n,\vkapp}(r)
 \Omega_{-\vkapp,m}(\omega)
 \end{array} \!\!\right)\!\!,
\end{align}
where relativistic quantum number $\vkapp =(l-j)(2j+1)$ and $\Omega_{-\vkapp,m}$
is spherical spinor. In these notations the radial integral for the magnetic
hyperfine constant for the point-like nuclear magnetic moment in the origin has the
form:
\begin{align}\label{HF_rad}
 I_{n',\vkapp',n,\vkapp} =
 \int_{0}^\infty\!\!
 \left(
 P_{n',\vkapp'}Q_{n,\vkapp}
 +Q_{n',\vkapp'}P_{n,\vkapp}
 \right)\frac{dr}{r^2}.
\end{align}

Magnetization of the nucleus is formed by the spin polarization of
nucleons and by the orbital motion of protons. \citet{BW50} noted
that if nuclear magnetization is localized at the spherical nuclear 
surface, then the spin contribution vanishes inside the nucleus, 
while the orbital one grows linearly from the center. Similar linear 
growth corresponds to the uniform spin distribution inside the nucleus.
Radial integral inside the nucleus of radius $R_N$ for this case has 
the form \cite{MP95}:
\begin{align}\label{HF_anom}
 I_{n',\vkapp',n,\vkapp}^\mathrm{nuc} =
 \int_{0}^{R_N}\!\!
 \left(
 P_{n',\vkapp'}Q_{n,\vkapp}
 +Q_{n',\vkapp'}P_{n,\vkapp}
 \right)\frac{r\, dr}{R_N^3}.
\end{align}
Outside the nucleus expression \eqref{HF_rad} still holds.

In our package we use the model of the uniformly charged ball and
inside the nucleus we use Taylor expansion for the radial functions
$P$ and $Q$:
\begin{align}\label{Taylor}
 P_{n,\vkapp}(r)|_{r\le R_N} = r^{|\vkapp|}
 \sum_{k=0}^M P_{n,\vkapp,k}\, x^k\,,
 \quad x=\frac{r}{R_N}\,.
\end{align}
With the help of this expansion we can calculate integral
\eqref{HF_anom} and nuclear contribution to integral \eqref{HF_rad}:
\begin{align}
 \nonumber
 &I_{n',\vkapp',n,\vkapp}^\mathrm{nuc}
 = R_N^{|\vkapp'|+|\vkapp|-1}
 \\
 \label{HF_Taylor}
 &\times
 \sum_{m=0}^M\sum_{k=0}^m
 \frac{P_{n',\vkapp',k}Q_{n,\vkapp,m-k}+Q_{n',\vkapp',k}P_{n,\vkapp,m-k}}
 {|\vkapp'|+|\vkapp|+m+2}\,,
 \\
 \nonumber
 &I_{n',\vkapp',n,\vkapp}^\mathrm{nuc,0}
 = R_N^{|\vkapp'|+|\vkapp|-1}
 \\
 \label{HF_Taylor0}
 &\times
 \sum_{m=0}^M\sum_{k=0}^m
 \frac{P_{n',\vkapp',k}Q_{n,\vkapp,m-k}+Q_{n',\vkapp',k}P_{n,\vkapp,m-k}}
 {|\vkapp'|+|\vkapp|+m-1}\,.
\end{align}

Using expression \eqref{HF_Taylor0} for two different nuclear radii
we can calculate charge correction to atomic HFS, while using
expression \eqref{HF_Taylor} we simultaneously account for charge
and magnetic corrections.

%
In order to disentangle these two corrections we introduce magnetic
radius of the nucleus $R_M$. We assume that expressions \eqref{HF_Taylor} 
and \eqref{HF_Taylor0} hold for $r\le R_M$ and $r>R_M$ respectively. 
For the volume distribution of magnetization these two expressions should 
match each other at $r=R_M$. However, for the surface distribution there 
may be a gap between them. We multiply \eqref{HF_Taylor} by a factor 
$(1-C_S)$ to account for this gap. Then $C_S=0$ gives smooth behavior at 
the surface and $C_S=1$ corresponds to the zero contribution of the volume
inside the nucleus. Our final expression for the
radial integral inside the nucleus combines integrand from Eq.\
\eqref{HF_anom} for $r\le R_M$ with the integrand from Eq.\
\eqref{HF_rad} for $R_M< r\le R_N$:
\begin{widetext}
\begin{align}
 \label{HFS_RM}
 I_{n',\vkapp',n,\vkapp}^\mathrm{nuc}(R_N,R_M)
 &=
 (1-C_S)\, I_{n',\vkapp',n,\vkapp}^\mathrm{nuc}(R_M)
 +\left(I_{n',\vkapp',n,\vkapp}^\mathrm{nuc,0}(R_N)
 -I_{n',\vkapp',n,\vkapp}^\mathrm{nuc,0}(R_M)\right),
 \\
 \label{HF_Taylor_RM}
 I_{n',\vkapp',n,\vkapp}^\mathrm{nuc}(R_M)
 &= R_M^{|\vkapp'|+|\vkapp|-1}
 \sum_{m=0}^M\sum_{k=0}^m
 \frac{P_{n',\vkapp',k}Q_{n,\vkapp,m-k}+Q_{n',\vkapp',k}P_{n,\vkapp,m-k}}
 {|\vkapp'|+|\vkapp|+m+2}
 \left(\!\frac{R_M}{R_N}\!\right)^m \!,
 \\
 \label{HF_Taylor0_RM}
 I_{n',\vkapp',n,\vkapp}^\mathrm{nuc,0}(R_M)
 &= R_M^{|\vkapp'|+|\vkapp|-1}
 \sum_{m=0}^M\sum_{k=0}^m
 \frac{P_{n',\vkapp',k}Q_{n,\vkapp,m-k}+Q_{n',\vkapp',k}P_{n,\vkapp,m-k}}
 {|\vkapp'|+|\vkapp|+m-1}
 \left(\!\frac{R_M}{R_N}\!\right)^m \!.
\end{align}
\end{widetext}

Equation \eqref{HFS_RM} describes several limiting cases. Taking
$R_M=0$ we return to the point magnetic dipole model. For $R_M=R_N$
and $C_S=0$ we get model \eqref{HF_Taylor}. Finally, for $R_M=R_N$
and $C_S=1$ we completely eliminate nuclear contribution to the
radial integral.

\subsection{Isotope effect for magnetic HFS}

Suppose we want to compare hyperfine constants $A_1$ and $A_2$ for
two isotopes with nuclear g factors $g_I^{(1)}$ and $g_I^{(2)}$,
nuclear charge radii $R^{(1)}_N$ and $R^{(2)}_N$, and magnetic radii
$R^{(1)}_M$ and $R^{(2)}_M$. We can write:
 \begin{align}
 \label{hfs_anom1}
 \frac{A_1}{A_2} = \frac{g_I^{(1)}}{g_I^{(2)}}
 \!\left(\!1
     -\lambda^C \frac{R^{(1)}_N-R^{(2)}_N}{R^{(1)}_N+R^{(2)}_N}
     -\lambda^M \frac{R^{(1)}_M-R^{(2)}_M}{R^{(1)}_M+R^{(2)}_M}
 \right)\!.
 \end{align}
The anomaly then has the following form:
 \begin{multline}
 \label{hfs_anom1a}
 ^{1}\Delta^{2} \equiv \frac{g_I^{(2)} A_1}{g_I^{(1)} A_2} -1 =
 \\
    = -\left(\lambda^C \frac{R^{(1)}_N-R^{(2)}_N}{R^{(1)}_N+R^{(2)}_N}
     +\lambda^M \frac{R^{(1)}_M-R^{(2)}_M}{R^{(1)}_M+R^{(2)}_M}
 \right)\,.
 \end{multline}
With the help of the method described above we can calculate
hyperfine constant for several values of $R_N$ and $R_M$. By solving
above equations for several radii, we can find $\lambda^C$ and
$\lambda^M$ and calculate the anomaly for the isotopes of interest.
Below we will see that parameters $\lambda^C$ and $\lambda^M$
themselves depend on the radii $R_N$ and $R_M$. Therefore it is be
better to use parameters $b_N$ and $b_M$ defined below (see
\Eref{derivation3}).

\subsection{Hydrogen-like ions}\label{Sec_H-like}

It is generally accepted that the observed hyperfine constant $A(R_N, R_M)$ of
a one-electron ion can be written in the following form:
 \begin{align}
 \label{hfs_Shabaev_1}
 A(R_N, R_M) = A_0(1-\delta(R_N))(1-\epsilon(R_M)).
 \end{align}
Here $A_0 \equiv A(0,0)$ is the factor, which is independent of
nuclear radii and $\delta(R_N)$ and $\epsilon(R_M)$ are the nuclear
charge distribution and magnetic distribution corrections
respectively. For a given $Z$ and electron state, they can be
written as:
 \begin{align}
 \label{H-scalings}
 \delta(R_N) = b_N R_N^{2\gamma -1},
 \qquad
 \epsilon(R_M) = b_M R_M^{2\gamma -1},
 \end{align}
where $b_N$ and $b_M$ are factors, which are independent of nuclear
radii, $\gamma = \sqrt{\vkapp^2 - (\alpha Z)^2}$, and $\alpha$ is
the fine structure constant. The expression for $A_0$ was obtained
in the analytical form as \cite{Sha94}:
 \begin{align}
 \label{hfs_Shabaev_2}
 A_0 = \frac{\alpha (\alpha Z)^3 g_I}{j(j+1)}
 \frac{m}{m_p}\frac{\vkapp(2\vkapp(\gamma +n_r) - N)}
 {N^4\gamma(4\gamma^2 -1)} mc^2.
 \end{align}
Here $m$ and $m_p$ are electron and proton masses, $g_I = \mu/I$ is
nuclear g factor, $j$ is the total electron angular momentum, $N=
\sqrt{n_{r}^2 + 2n_r\gamma + \vkapp^2}$, $n_r$ is radial quantum
number.

It follows from Eqs.\ \eqref{hfs_Shabaev_1} and \eqref{H-scalings},
that if we calculate HFS constant numerically for different $R_N$
and $R_M$, we should get following dependence on the radii:
 \begin{align}
 \label{hfs_fit_1}
 A(R_N, R_M) = A_0(1 - b_{N} R_N^{2\gamma -1})(1 - b_{M} R_M^{2\gamma -1}).
 \end{align}
This expression defines the dependence of parameters $\lambda^C$ and
$\lambda^M$ from \eqref{hfs_anom1} on the radii $R_N$ and $R_M$. For
example, from one hand, we have:
 \begin{align}
 \label{derivation1}
 \frac{A(R_{N}+\rho,R_{M})}{A(R_{N}-\rho,R_{M})} = 1-\lambda^C(R_{N}) \frac{\rho}{R_{N}}.
 \end{align}
From the other hand:
 \begin{align}\label{derivation2}
 \frac{A(R_{N}+\rho,R_{M})}{A(R_{N}-\rho,R_{M})}
 = 1 + 2\rho\frac{\partial A(R_{N},R_{M})/\partial R_{N}}{A(R_{N},R_{M})}.
 \end{align}
Then, from Eq.\ \eqref{derivation1} we get:
\begin{multline}\label{derivation3}
    \lambda^C(R_N) \approx \frac{2(2\gamma -1) b_{N} R_N^{2\gamma -1}}
    {1 - b_{N} R_N^{2\gamma -1}}
 \\
    \approx 2(2\gamma -1) b_N R_N^{2\gamma -1}.
\end{multline}
Similar expressions can be obtained for $\lambda^M(R_M)$.

For the point-like magnetic dipole approximation ($R_M=0$) the
magnetic correction $\epsilon$ is equal 0, and the hyperfine
constant can be fitted by the function:
 \begin{align}
 \label{A_PD_fit}
 A (R_N, 0) = A_0(1 - b_{N} R_N^{2\gamma -1}).
 \end{align}
For the uniform distribution of the charge and magnetic moment with
$R_N = R_M$ we get:
 \begin{align}
 \label{A_UD_fit}
 A (R_N, R_N) = A_0(1 - (b_N + b_M) R_N^{2\gamma -1})
 \end{align}

\subsection{Many-electron atoms}\label{Sec_many_electron}

Since the one-electron radial integrals are defined, we can
calculate atomic HFS using many-electron wave functions and account
for electronic correlations as described in Ref.\ \cite{DFKP98}.
Using Eqs.\ (\ref{HFS_RM} -- \ref{HF_Taylor0_RM}) we can calculate
atomic HFS constants for arbitrary radii $R_N$ and $R_M$ with the
only constraint that $R_N \ge R_M$.
We can do configuration interaction calculations with the frozen core and few valence electrons. 
Then we can add core-valence correlation corrections with
the help of the many-body perturbation theory. On this stage we
substitute valence radial integrals with the effective ones, which
account for the spin polarization of the core. The latter are
obtained by solving random-phase approximation (RPA) equations.

Effective radial integrals may have significantly different
dependence on the parameters of the nucleus, than initial ``bare''
integrals. This is particularly true for the orbitals with high
angular momentum. Because of the centrifugal barrier these orbitals
do not penetrate inside the nucleus and bare radial integrals do not
depend on the nuclear size. On the other hand, spin-polarization of
the core always include polarization of the core $s$ and $p_{1/2}$
shells. Because of that all effective radial integrals are sensitive
to the nuclear charge and magnetic distributions.

In general, we can divide all correlation corrections in two
classes: corrections, which mix orbitals within one partial wave,
and the ones which mix different partial waves. For example, the
self-energy type corrections belong to the fist class. They mix core
and valence orbitals of the same symmetry and can significantly
change the orbital density at the origin. Therefore, these
corrections change the size of the HFS matrix elements. On the other
hand, all orbitals of the same symmetry have practically the same
sensitivity to the nuclear distributions. Thus, such correlation
corrections do not affect parameters $b_N$ and $b_M$ and the HFS
anomaly \eqref{hfs_anom1a}. RPA corrections belong to the second
class, which significantly contribute to the HFS anomaly.

\section{Results and discussion}

\subsection{HFS anomaly for H-like thallium ion}

\begin{figure}[htb!]
     \centering
     \includegraphics[height=7cm]{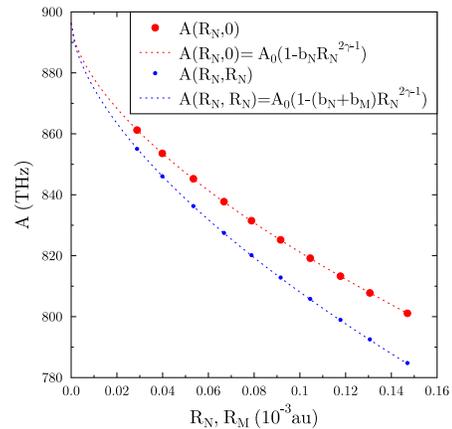}
     \caption{The dependence of the HFS constant $A(R_N, R_M)$ for the ground state of H-like
     Tl ion from nuclear charge and magnetic radii. Dots and circles correspond to
     the computed values. Dashed lines correspond to the fits by Eqs.\ \eqref{A_PD_fit} and
     \eqref{A_UD_fit}.
     \label{frg:A_fit}}
 \end{figure}

In this section we calculate HFS constants of the $1s$, $2s$, and
$2p_{1/2}$ states of Tl$^{80+}$ for different radii $R_N$ and $R_M$
and compare our results with analytical expressions from Ref.\
\cite{Sha94}. Figure \ref{frg:A_fit} shows the dependence of the
hyperfine constant $A(1s)$ on the radii $R_N$ and $R_M$. We see very
good agreement with Eqs.\ \eqref{A_PD_fit} and \eqref{A_UD_fit}.

 \begin{table}[!htb]
\caption{\label{tbl:h-like} 
Compilation of the fitting parameters for HFS of H-like Tl ion: 
$A_0$ is HFS constant for point-like nucleus, $\delta$ and $\epsilon$ are the nuclear charge and
magnetization distribution corrections parametrized by $b_N$ and $b_M$ coefficients respectively. 
We use g factor $g_I = 3.27640$. Corrections $\delta$ and $\epsilon$ for $\rm ^{203}Tl$ are calculated for $R_N=R_M=0.1306 \times 10^{-3}$ au.}
\begin{tabular}{lcccc}
\hline \hline
&&$1s$&$2s$& $2p_{1/2}$ \\
\hline
$A_0$ (THz)                                  &fit.& 896.4  & 144.9& 45.0 \\
                         &Eq.~\eqref{hfs_Shabaev_2}& 895.7  &144.8 & 45.0 \\
$b_{N}$                                      &fit.& 0.3441 & 0.3671& 0.0960 \\
$\delta$ for $\rm ^{203}Tl^{80+}$            &fit.& 0.0988 & 0.105 & 0.028\\
                                             &Ref.\ \cite{STKAY97}& 0.0988& -- & --\\
$b_{M}$                                      &fit.& 0.0599 & 0.0638& 0.0176\\
$\epsilon$ for $\rm ^{203}Tl^{80+}$          &fit.&0.0172  &0.0183& 0.0051\\
                                             &Ref.\ \cite{STKAY97}&  0.0179& -- & --\\
\hline \hline
\end{tabular}
\end{table}

Table~\ref{tbl:h-like} summarizes our results for H-like Tl ion. We
see perfect agreement of the calculated and analytical values of
$A_0$ for all three states. Charge and magnetic corrections $\delta$
and $\epsilon$ were calculated in Ref.\ \cite{STKAY97} for the $1s$
state of the isotope $^{203}$Tl. These analytical values are also in
good agreement with our numerical results.

 \begin{figure}[htb!]
     \centering
     \includegraphics[height=7cm]{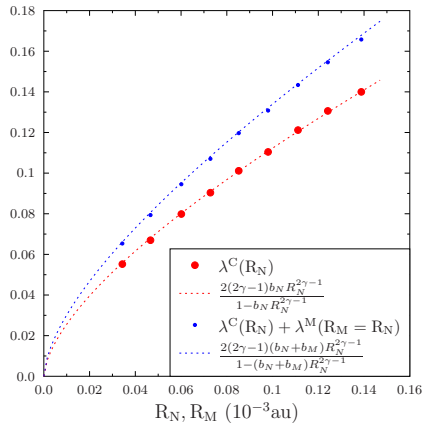}
     \caption{Dependence of the parameters $\lambda^C(R_N)$ and $\lambda^M(R_M)$
      (see Eq.~\eqref{hfs_anom1}) on the charge and magnetic radii of the nucleus
      for the ground state of H-like Tl ion. Computed values represented by points.
      The curves correspond to the fits with Eq.~\eqref{derivation3}.
      \label{fgr:lambda_H-like}}
 \end{figure}

Figure \ref{fgr:lambda_H-like} shows how parameters $\lambda$ for
the $1s$ state depend on the radii $R_N$ and $R_M$. On one hand, we
see perfect agreement with the analytical expression
\eqref{derivation3}. On the other hand, it means that these
parameters strongly depend on the nuclear size. Because of that they
can not be treated as constant even for the isotopes with similar
radii. Therefore it is better to use parameters $b_N$ and $b_M$
defined by \Eref{hfs_fit_1}.

According to our calculations (see Table \ref{tbl:h-like}) the
ratios of the parameters $b_N$ and $b_M$ for $1s$ and $2s$ states
are close to unity: $\frac{b_N(1s)}{b_N(2s)} = 0.937$ and $\frac{b_M
(1s)}{b_M (2s)} = 0.939$. This is expected, as wave functions of the
same symmetry should be proportional to each other inside the
nucleus. Similar ratios for $1s$ and $2p_{1/2}$ states are
$\frac{b_N (1s)}{b_N (2p_{1/2})} = 3.58$ and $\frac{b_M (1s)}{b_M
(2p_{1/2})} = 3.40$. Again, one can expect that these ratios only
weakly depend on the principle quantum numbers.

\subsection{HFS anomaly of neutral thallium atom}

The ground configuration of the neutral thallium is $[1s^2 \dots
6s^2]6p$ and the ground multiplet includes two levels, $6p_{1/2}$
and $6p_{3/2}$. The lowest level of the opposite parity is $7s$.
Most of the experiments and calculations of the HFS in neutral
thallium deal with these three levels. If we treat thallium as a
one-electron system with the frozen core $[1s^2\dots 6s^2]$, we can
do calculation using Dirac-Hartree-Fock (DHF) method. In this case
the dependence of the HFS constants on the nuclear radii is similar
to the one-electron ion.

\begin{table}[tbh]
\caption{\label{tbl_tl} Compilation of the fitting parameters for
HFS of neutral Tl atom: $A_0$ is HFS constant for point-like
nucleus, $\delta$ and $\epsilon$ are the nuclear charge and
magnetization distribution corrections parametrized by $b_N$ and
$b_M$ coefficients respectively. We use g factor $g_I = 3.27640$.
Corrections $\delta$ and $\epsilon$ for $\rm ^{203}Tl$ are
calculated for $R_N=R_M=0.1306 \times 10^{-3}$ au. Calculations are
done within DHF and DHF+RPA approximations.}
\begin{tabular}{lcccccc}
\hline\hline
&\multicolumn{3}{c}{DHF}&\multicolumn{3}{c}{DHF+RPA} \\
&$6p_{1/2}$&$7s$&$6p_{3/2}$&$6p_{1/2}$&$7s$&$6p_{3/2}$ \\
\hline
$A_0$ (GHz)                                  & 18.308&  8.942&1.315& 22.960& 12.586& -2.423\\
$b_{N}$                                      & 0.1054& 0.3709&0    & 0.1352& 0.3517& 0.5302\\
$\delta$ for $\rm ^{203}Tl$                  & 0.0303& 0.1064&0    & 0.0388& 0.1009& 0.1522\\
$b_{M}$                                      & 0.0195& 0.0621&0    & 0.0250& 0.0643& 0.0989\\
$\epsilon$ for $\rm ^{203}Tl$                & 0.0056& 0.0178&0    & 0.0072& 0.0185& 0.0284\\
\hline\hline
\end{tabular}
\end{table}

\begin{figure}[tbh]
     \centering
     \includegraphics[height=7cm]{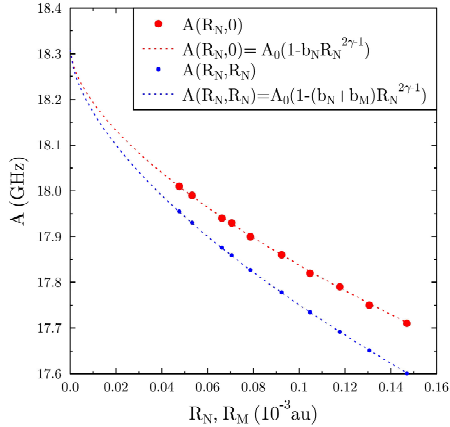}
     \caption{The dependence of the HFS constant $A(R_N, R_M)$ for the ground state of
     neutral Tl ion from nuclear charge and magnetic radii. Dots and circles correspond to
     the computed values within Dirac-Hartree-Fock method. Dashed lines correspond to the
     fits by Eqs.\ \eqref{A_PD_fit} and
     \eqref{A_UD_fit}.
\label{fgr:Tl}}
 \end{figure}

In DHF approximation the HFS constant $A(6p_{3/2})= 1.30$ GHz is
very small and practically does not depend on $R_N$ and $R_M$ (see
Table \ref{tbl_tl}). At the same time, the HFS constants
$A(6p_{1/2})$ and $A(7s)$ are well described by Eqs.\
(\ref{A_PD_fit},~\ref{A_UD_fit}) (see Fig.\ \ref{fgr:Tl}). 
According to our calculations, the ratios between coefficients $b_N$ and $b_M$ for $s$ and $p_{1/2}$ waves are close to the respective ratios in H-like ion. 
For example, the ratios of these constants for $1s$ state of the ion and $7s$ state of the neutral atom are $\frac{b_N(1s)}{b_N(7s)} = 0.928$ and $\frac{b_M(1s)}{b_M(7s)} = 0.965$.
This result is compatible with assertion that the hyperfine anomaly measured for the $s$ states in Rb is independent of the principal quantum number~\cite{GZO07}.
Atomic ratios for $7s$ and $6p_{1/2}$ are: $\frac{b_N
(7s)}{b_N (6p_{1/2})} = 3.52$ and $\frac{b_M(7s)}{b_M(6p_{1/2})} = 3.18$, while for the H-like ion we had 3.58 and 3.40 respectively.

Situation changes when we include spin-polarization of the core via
RPA corrections. These corrections mix partial waves and the state
$6p_{3/2}$ partly acquire $s$ and $p_{1/2}$ character. This leads to
significant change of the size and even the sign of the constant
$A(6p_{3/2})$. At the same time this constant becomes very sensitive
to the distributions of charge and magnetic moment inside the
nucleus.
RPA corrections for the $7s$ and $6p_{1/2}$ states are smaller than
for $6p_{3/2}$, but also significant. They lead to effective mixing
of the $s$ and $p$ waves. Because of that the ratios of the
respective coefficients decrease a little, but are still much bigger
than unity:
 \begin{align}
 \label{fin_ratio}
 \frac{b_N(7s)}{b_N(6p_{1/2})} = 2.60\,,
 \qquad
 \frac{b_M(7s)}{b_M(6p_{1/2})} = 2.57\,.
 \end{align}
We conclude that in the DHF+RPA approximation, the anomaly for the
$7s$ state is still significantly stronger, than for $6p_{1/2}$
state. The anomaly for the $6p_{3/2}$, on the contrary, becomes the
largest. This conclusion holds when we include more correlation
corrections, as it was done in \cite{DFKP98}.

Using experimentally measured value for HFS anomaly \eqref{hfs_anom1} for
the ground state $6p_{1/2}$ of the thallium two stable isotopes
$^{205}\Delta^{203}(6p_{1/2})=-1.036(3) \times 10^{-4}$ \cite{LP56},
and the ratios \eqref{fin_ratio} calculated here, we can obtain
corresponding value for the $7s$ state within $R_N = R_M$ approximation:
$^{205}\Delta^{203}(7s)=-2.7 \times 10^{-4}$.
This value is significantly lower, than experimental value
$-4.7(1.5)\times 10^{-4}$ obtained in Ref.\ \cite{RLM00}.

\section{Nuclear magnetization}

\begin{table}[tbh]
\caption{\label{spin_part} Magnetic HFS constants (MHz) for $\rm
^{203}Tl$ calculated for $R_N= 0.1306 \times 10^{-3}$ a.u.\ and
different values of $R_M$ and $C_S$.}
\begin{tabular}{crrrrr}
\hline\hline
$R_M/R_N$
&\multicolumn{1}{c}{0}
&\multicolumn{1}{c}{1}
&\multicolumn{1}{c}{1}
&\multicolumn{1}{c}{0.9}
&\multicolumn{1}{c}{0.8}
\\
$C_S$
&\multicolumn{1}{c}{0}
&\multicolumn{1}{c}{1}
&\multicolumn{1}{c}{0}
&\multicolumn{1}{c}{0.345}
&\multicolumn{1}{c}{0.805}
\\
\hline
&\multicolumn{5}{c}{DHF}\\
$A(6p_{1/2})$  &  17754.26 &  17590.89 &  17650.92 &  17650.83 & 17650.91 \\
$A(6p_{3/2})$  &   1314.50 &   1314.50 &   1314.50 &   1314.50 &  1314.50 \\
$A(7s_{1/2})$  &   7990.45 &   7732.08 &   7826.81 &   7826.64 &  7826.71 \\
$A(7p_{1/2})$  &   1970.07 &   1951.94 &   1958.60 &   1958.59 &  1958.60 \\
$A(7p_{3/2})$  &    188.10 &    188.10 &    188.10 &    188.10 &   188.10 \\
\hline
&\multicolumn{5}{c}{DHF+RPA}\\
$A(6p_{1/2})$  &  22068.38 &  21806.94 &  21903.16 &  21903.01 &  21903.16 \\
$A(6p_{3/2})$  &$ -2057.68$&$ -1949.05$&$ -1989.17$&$ -1989.12$&$ -1989.20$\\
$A(7s_{1/2})$  &  11322.86 &  10957.33 &  11091.66 &  11091.44 &  11091.60 \\
$A(7p_{1/2})$  &   2029.95 &   2014.27 &   2020.03 &   2020.02 &   2020.02 \\
$A(7p_{3/2})$  &    112.78 &    115.35 &    114.39 &    114.39 &    114.39 \\
\hline\hline
\end{tabular}
\end{table}

In this section we discuss how much we can say about nuclear
magnetization from the atomic hyperfine structure measurements. In
the model we use here this magnetization is described by magnetic
radius $R_M$ and additional parameter $C_S$ \eqref{HFS_RM}. \Tref{spin_part}
presents results of HFS calculations in DHF and DHF+RPA
approximations for $^{203}$Tl with different values of these
parameters. Charge radius in all calculations is taken to be
$R_N=0.1306 \times 10^{-3}$~a.u. The two limiting cases are given
by $R_M=0$ and $R_M=R_N,\,C_S=1$, which correspond to the largest
and the zero nuclear contribution to the HFS radial integrals. All
other results lie between these ones for both approximations. The
last three columns in \Tref{spin_part} correspond to three different
values of $R_M$. Nuclear contribution grows when we decrease
magnetic radius $R_M$ and decreases with increasing parameter
$C_S$. For each magnetic radius we choose $C_S$ so that all five HFS
constants remain constant for both approximations (!). It is
particularly important because nuclear contributions for DHF and RPA
approximations are very different. We can conclude that already our
simple model of nuclear magnetization is degenerate and nuclear
parameters $R_M$ and $C_S$ can not be uniquely found from atomic
HFS. Consequently, there is no point in using more complex nuclear
models.

\section{Conclusions}

In this work we propose a method for calculation hyperfine structure
constants of many-electron atoms as functions of nuclear charge and
magnetic radii $R_N$ and $R_M$. The HFS anomaly in this method can
be parametrized by $b_N$ and $b_M$ coefficients. If HFS anomaly is
known from the experiment, then we can use coefficients $b_N$ and
$b_M$ to determine these radii. Alternatively, we can use these
coefficients to improve the accuracy for nuclear g factors of the
short lived isotopes, obtained from the ratios of the HFS constants.
We tested this method by calculating HFS constants of H-like thallium
ion and obtained fairly good agreement with analytical expressions from
Refs.\ \cite{Sha94,STKAY97}. Then we made calculations for neutral
thallium atom described as a one-electron system. In the
Dirac-Hartree-Fock approximation the ratios between hyperfine
anomalies of $s$ and $p_{1/2}$ states of neutral Tl atom and
respective H-like ion are the same. However when we include
spin-polarization of the core via RPA corrections, only the
hyperfine anomaly for the $7s$ state remains stable. The ratios
between $7s$ and $6p_{1/2}$ states change by roughly 30\%, and the
anomaly for the $6p_{3/2}$ state becomes very large. We conclude,
that for the precision measurements of g factors it is preferable to
use the hyperfine constants for $s$ states, while the $p_{3/2}$
states are least useful.

\acknowledgments
Thanks are due to Prof. Vladimir M. Shabaev, Prof. Ilya I. Tupitsyn
and Dr. Leonid V. Skripnikov for helpful discussions. The work was
supported by  the Russian Foundation for Basic Research (grant \#
17-02-00216).


\end{document}